\documentstyle[12pt]{article}

\newcommand{\disp}{\displaystyle}
\begin{document}
\thispagestyle{empty}

\begin{center}
{\Large\bf Bosonization for 2D Interacting Fermion Systems:
Non-Fermi Liquid Behavior}
\end{center}
\bigskip
\begin{center}
Y. Yu, Y. M. Li and N. d'Ambrumenil \\
Department of Physics, University of Warwick,\\
Coventry, CV4 7AL, U. K. \\
\end{center}
\bigskip
\bigskip
\begin{center}
{\it ABSTRACT}
\end{center}
{\it
Non-Fermi liquid behavior is found
for the first time in a two-dimensional (2D) system
with non-singular interactions
using Haldane's bosonization scheme.
The bosonized system
is solved exactly by a generalized
Bogoliubov transformation.
The fermion momentum distribution, calculated using a
generalized Mattis-Lieb technique,  exhibits
a non-universal power law in the vicinity of
the Fermi surface for intermediate interaction strengths.}
\eject

In recent years, there has been great interest in possible
non-Fermi-liquid-like
behavior in the ground state of two-dimensional (2D) strongly correlated
fermionic systems [1]. It is well-known that in one dimension
the ground state of a fermionic system with arbitrarily weak
interactions is not a Fermi liquid but a state usually known
as a Luttinger liquid
[2]. For the 2D
case, many studies have suggested  that, in the limit of weak
interactions, this behavior does not appear [3]. Studies of
systems with a small number of particles also suggest Fermi liquid behavior in
2D [4]. Although Khveshchenko {\it et al}  have found non-Fermi-liquid
behavior in 2D, they worked only  with a singular
long-range current-current
interaction [5].

The one-dimensional Luttinger liquid solution can be obtained by a
bosonization procedure [2]. Haldane has recently generalized this
procedure to solve 2D models [6] and his method has been further
developed in [5]. In this paper, we show using this procedure
that, even  for non-singular short-range interactions, 2D systems may show
non-Fermi-liquid behavior if the interaction is strong enough.
However, in the limit of weak interactions
it is difficult to distinguish between a
Fermi liquid and a non-Fermi liquid.

We diagonalize the bosonized model
exactly using a generalized Bogoliubov transformation.
The fermion momentum distribution $n({\bf p})$ at zero temperature
is then calculated using
a generalization to higher dimensions of the method used by Mattis
and Lieb [7] in 1D.
The power law obtained in 1D for $n(p)$ near $p_F$, $n(p)\sim (p-p_F)^
\delta$, also emerges in 2D. The exponents, $\delta$,
can be calculated numerically.
We look in particular at the case of a $\delta$-function
interaction and find that the
variation of $\delta$ with coupling constant is similar to that found in 1D,
although $\delta$ is always smaller in 2D  than
1D for comparable values of the coupling constants.

We begin with a brief description of  Haldane's procedure.
Consider the fermionic system with the spectrum $\epsilon({\bf p})
=v(|p_x|+|p_y|)$. The spectrum has 4 different
branches with one in each quadrant
of the ${\bf p}$-plane. One constructs 4
different density operators $B_\alpha({\bf q})=\sum c^\dagger_{
\alpha,{\bf p+q}}
c_{\alpha{\bf
p}}$ where $\alpha$ is the quadrant index and ${\bf p}$ and ${\bf p+q}$
lie in the $\alpha$-th quadrant. It is easy to show that those operators obey
the commutation relations
$$
[B_\alpha({\bf q}), B_\beta({\bf
-q}')]=-\delta_{\alpha\beta}
\delta_{\bf q,q'}\frac{\Lambda\Omega}{(2\pi)^2}{\bf n}_\alpha\cdot{\bf q},
\eqno(1)
$$
where
$\Lambda$ and ${\bf n}_\alpha$  are  the length and normal vector of the Fermi
surface in the $\alpha$-th
quadrant. $\Omega$ denotes the volume of the system.
In terms
of the commutation relations between the kinetic energy and $B_\alpha({\bf
q})$,
the total bosonized Hamiltonian reads
$$
H=\frac{(2\pi)^2v}{\Lambda\Omega}\sum_{{\bf q},\alpha,\beta}(
\delta_{\alpha\beta}+\frac{\Lambda
\Gamma_{\alpha\beta}({\bf q})}{v})B_\alpha({\bf q})B_\beta(-{\bf q}),\eqno(2)
$$
if the fermionic interaction has the form $ \frac{1}{\Omega}
\sum_{\bf p,p',q}\Gamma({\bf p,p',q})c^\dagger _{\bf p+q}
c^\dagger_{\bf p'-q}c_{{\bf p}'}c_{\bf p}$.
Since the definition of the $B_\alpha({\bf q})$ excludes contribution to
the density operators in which ${\bf p}$ and ${\bf p}+{\bf q}$ are in
different quadrants, certain scattering terms are not included in the
bosonized Hamiltonian. These describe the so-called around-the-corner
(ATC) scattering. In the limit $q\to 0$ the contribution of ATC
scattering vanishes as $q/\Lambda$. We will return to this question later.

Haldane has pointed out that the 2D system described by (2) will converge
to the Fermi liquid fixed point in the limit $\Lambda\to 0$
for non-singular interactions. Haldane's observation is obviously
correct when the summation over $\alpha,\beta$ is
over four quadrants.
%Therefore some singular interactions are invoked to obtain
% non-Fermi-liquid behaviors.
However,
if the summation over $\alpha$ and $\beta$ is taken over an infinite
number of sectors instead of four quadrants, as $\Lambda
\rightarrow 0$, the situation may change as we
shall see below.

We consider an interacting fermion model described by the Hamiltonian
$$
H=\sum_{{\bf p},s}\epsilon({\bf p})c^\dagger_{{\bf p}s}c_{{\bf p}s}
+\frac{(2\pi)^2}{\Omega}\sum_{\bf q}V(q)\sum_{{\bf p,p'},s,s'}
c^\dagger_{{\bf p+q},s}c^\dagger_{{\bf p'-q},s'}c_{{\bf p}'s'}
c_{{\bf p}s}, \eqno(3)
$$
where Planck's constant $\hbar$, the speed of light $c$  and lattice
length $a$ have been set to unity.
We assume the spectrum $\epsilon({\bf p})
=v(|{\bf p}|-p_F)$ where $p_F$ is the Fermi momentum. Thus the
Fermi surface is a circle with radius $p_F$.
Following Khveshchenko {\it et
al} [5], we divide this circular Fermi surface into an
$M$-sided polygon ($M\to\infty$), whose
side length is $\Lambda$ with $\Lambda
\ll p_F$. The whole momentum space is thus divided into $M$ sectors.
The $\alpha$-th sector, $T_\alpha$, is defined by the direction
of the unit vector ${\bf n}_\alpha$.
In order to understand the low energy behavior of the system,
we only consider a shell of width  $L$ ($L\sim\Lambda$) around
the Fermi surface.
The bosonic operators are defined as
$B_{\alpha s}({\bf q}) =\sum_{{\bf p}\in \Lambda_\alpha,{\bf p+q}\in
T_\alpha}c^\dagger_
{\alpha,{\bf p+q},s}c_{\alpha {\bf p} s}$, where $\alpha $ runs from 0 to
$M-1$.
$\Lambda_\alpha$ is the area $L\times\Lambda$
and centered at ${\bf p}^{(\alpha)}_F$, the Fermi momentum in the $\alpha$-th
direction.
The bosonic operators
satisfy the commutation relations of Eq.(1).

The kinetic energy can be written in terms of the
$B_{\alpha s}({\bf q})$ [5]
$$
H_{kin}\approx \frac{(2\pi)^2v}{\Lambda\Omega}\sum_{{\bf q}, \alpha,s}
B_{\alpha s}({\bf q})B_{\alpha s}(-{\bf q})+O(q^2).\eqno(4a)
$$
The interaction is bosonized to give
$$ \begin{array}{rcl}\disp
H_{V}&=&\disp\frac{(2\pi)^2}{\Omega}\sum_{{\bf q},\alpha,s}V'(q) B_{\alpha s}
({\bf q}) B_{\alpha s}(-{\bf q})+V(q) B_{\alpha s}({\bf q})
B_{\alpha,-s}(-{\bf q})\\[3mm]\disp
&+&\disp\frac{(2\pi)^2}{\Omega}\sum_{s,s',\alpha\not=\beta}V({\bf q})
B_{\alpha s}({\bf q})B_{\beta s'}(-{\bf q}).
\end{array}\eqno(4b)
$$
Here $V'(q)=V(q)-V_1(q)$ with $V(q)$ corresponding to the
contribution from direct scattering
and $V_1(q)$ to that from  exchange scattering. For a $\delta$-function
interaction, $V'(q)$ vanishes, as required by Pauli's principle [8].
In Eq.(3a), we have restricted the momenta ${\bf p}$
and ${\bf p+q}$ included in the interaction term
of (3) to $\Lambda_\alpha$ and $T_\alpha$, respectively,
as done in refs.[6,5].

We now solve the bosonized model described by $H=H_{kin}+H_V$.
The commutation
relations for the
$B_{\alpha s}({\bf q})$ allow creation and annihilation operators
to be defined as follows. For ${\bf n}_\alpha\cdot{\bf q}>0$,
$$\begin{array}{l}
B_{\alpha s}({\bf q})=\disp\sqrt{\frac{\Lambda\Omega}{(2\pi)^2}}\omega^{1/2}
_{q,\alpha} a^\dagger_{\alpha s}(-{\bf q}),\\[3mm]
B_{\alpha s}(-{\bf q})=\disp\sqrt{\frac{\Lambda\Omega}{(2\pi)^2}}\omega^{1/2}
_{q,\alpha} a_{\alpha s}(-{\bf q}),
\end{array}\eqno(5)
$$
where $\omega_{q,\alpha}=|{\bf n}_\alpha\cdot{\bf q}|$.
For ${\bf n}_\alpha\cdot{\bf q}<0$, $a^\dagger_{\alpha s}(-{\bf q})$
and $a_{\alpha s}(-{\bf q})$ are replaced by $a_{\alpha s}({\bf q})$ and
$a^\dagger_{\alpha s}({\bf q})$ respectively. The operators $a_{\alpha s}
(\pm{\bf q})$ obey
the canonical commutation relations $[a_{\alpha s}(\pm{\bf q}),
a^\dagger_{\beta s'}(\pm{\bf q}')] = \delta_{\alpha\beta} \delta_{\bf q,q'}
\delta_{ss'}$.
Again we divide the {\bf q}-plane into $M$ regions. Each of them
is the region between ${\bf n}_\alpha$ and ${\bf n}_{\alpha+1}$,
and is denoted by $D_\alpha$. Without losing generality, we consider
$M=4N$ for convenience.
Using $\omega_{{\bf q},\alpha}=\omega_{{\bf q},\alpha+2N}=\omega_{{\bf q},
\alpha+4N}$, the
bosonized Hamiltonian can now be rewritten as
$$
H=2\sum_{n=0}^{2N-1}\sum_{{\bf q}\in D_n}\sum_{s,s'}
{\cal A}^\dagger_{n,s,{\bf q}}
\Omega_{s,s',{\bf q}}{\cal A}_{n,s',{\bf q}}.\eqno(6)
$$
For computational convenience, we organize the $4N$-component vectors,
${\cal A}^\dagger_{n,s}$,
so that the first $N$ and last $N$ components are the
creation operators.
For example, we would have for the cases $n=0$ and $n=1$:
$$ \begin{array}{l}
{\cal A}^\dagger_{0,s}=[a^\dagger_{0s},...,a^\dagger
_{N-1,s},a_{Ns},...,a_{3N-1,s},
a^\dagger_{3N,s},...,a^\dagger_{4N-1,s}],\\[3mm]
{\cal A}^\dagger_{1,s}=[a^\dagger_{0,s},...,a^\dagger_{N-1,s}
,a_{3N,s},a_{N+1,s},...,
a_{3n-1,s},a^\dagger_{Ns},a^\dagger_{3N+1,s}
,...,a^\dagger_{4N-1,s}],
\end{array}\eqno(7)
$$
and so on.
Here $a^\dagger_{\alpha,s}\equiv a^\dagger_{\alpha,s}(-{\bf q}_n)$
and $a_{\alpha,s}\equiv a_{\alpha,s}({\bf q}_n)$ and ${\bf
q}_n$ denotes ${\bf q}\in D_n$.
The elements of the matrix
$\Omega_{s,s'}$ are given by
$$
\begin{array}{l}
\Omega_{\alpha\beta,ss'}({\bf q})=[(v+\Lambda V'(q))\delta_{ss'}+
\Lambda V(q)\delta_{s,-s'}]\omega_{{\bf q},\alpha},~~~~~
{\rm for}~~\alpha=\beta,
\\[2mm]
\Omega_{\alpha\beta,ss'}({\bf q})=\Lambda V(q)\omega^{1/2}_{{\bf q},\alpha}
\omega_{{\bf q},\beta}^{1/2},~~~~~~~~~~~~~~~~~~~~~~~~~~~~~~~~{\rm for}
{}~~\alpha\not=\beta.
\end{array}
$$

We introduce the charge density operators, $R_{\alpha,{\bf q}}$, and
spin density operators, $S_{\alpha,{\bf q}}$, via
the relation
$ a_\alpha(\pm{\bf q})=
\frac{1}{\sqrt{2}}(R_{\alpha {\bf q}}+s~S_{\alpha{\bf q}})$.
The Hamiltonian (6) can therefore be expressed as
$H=H_\sigma+H_\rho$,
$$\begin{array}{l}\disp
H_\sigma=\sum_{n,D_n,\alpha} S^\dagger_{\alpha,{\bf q}_n}
\tilde\Omega^\sigma
_{\alpha\alpha,{\bf q}_n} S_{\alpha,{\bf q}_n},\\[2mm]\disp
H_\rho=\sum_{n,D_n,\alpha,\beta}\Gamma^\dagger_{\alpha,{\bf q}_n}
\tilde\Omega^\rho_{\alpha\beta,{\bf q}_n}\Gamma_{\beta,{\bf q}_n},
\end{array}\eqno(8)
$$
where
$\tilde\Omega^{\rho,\sigma}_{\alpha\alpha,{\bf,q}}
=(v+\Lambda V'(q)\pm\Lambda V(q))\omega_{\alpha,
{\bf q}}$ with $\rho$ corresponding to `+' and $\sigma$ to `$-$', and
$\tilde\Omega^\rho_{\alpha\beta,{\bf q}}=2
\Omega_{\alpha\beta,ss'}$ for $\alpha\not=\beta$.
The vectors $\Gamma$ correspond to vectors ${\cal A}$ in (7)
with $a$'s replaecd by $R$'s.

The spin density part of the Hamiltonian, $H_\sigma$, is already diagonal.
There are $4N$ excitation branches  with energies $\tilde\Omega^
\sigma_{\alpha\alpha,{\bf q}}$
. The branches $S_{\alpha,{\bf q}}$ and $S_{\alpha +2N,{\bf q}}$ are
degenerate.

We need to diagonalize
the density part of the Hamiltonian, $H_\rho$.
Formally, this can be done
by setting up the equation
$$
f(\lambda)={\rm Det} |\tilde\Omega^\rho_{aN+i,bN+j}-\Delta_a\lambda
\delta_{ab}\delta_{ij}|=0,\eqno(9)
$$
where $i,j=1,..,N$, and $a,b=0,...,3$. With our ordering
of the $4N$-components of the vector ${\cal A}$, $\Delta _a
=+1$ for $a=0,3$ and $\Delta_a=-1$ for $a=1,2$. Since
$\omega_{\alpha, {\bf q}}=\omega_{\alpha+2N,{\bf q}}$, we have
$f(\lambda)=f(-\lambda)$.
The $4N$-component `extended'
eigenvectors, $X^{(\mu)}$, satisfy:
$$
\Omega^\rho X^{(\mu)}=\lambda_\mu\Delta
X^{(\mu)},                                                  \eqno(10)
$$
for each $\lambda_\mu$.
Here the matrix $\Delta=(\Delta
_a\delta_{ab}
\delta_{ij})$.
For real $\lambda_\mu$ [9],  we have the
extended normalization and orthogonality relations
$$
X^{-1^T}\Delta X^{-1}=\Delta',
\eqno(11)
$$
where $X^{-1}=(X^{(0)},...,X^{(4N-1)})$
is the inverse of the transformation matrix $X$ and $\Delta'=(\Delta_\mu
\delta_{\mu\mu'})$ with $\Delta_\mu$ equal +1 for $\lambda_\mu>0$
and $-1$ for $\lambda_\mu<0$. The transformation
matrix, $X=\Delta'X^{-1^T}\Delta$, can be easily obtained
from the orthogonality relations (11).
Using Eqs.(10) and (11), $H_\rho$ is diagonalized to $H_\rho=\sum|\lambda_\mu
({\bf q})|\theta^\dagger_{\mu,{\bf q}_n}\theta_{\mu,{\bf q}_n}$,
where the $\theta_{\mu,{\bf q}_n}^\dagger$ and
$\theta_{\mu,{\bf q}_n}$ are true bosonic creation and annihilation
operators.

To determine the nature of the ground state of the Hamiltonian (8) and,
in particular, whether it is a Fermi liquid or not we look at the momentum
distribution, $n({\bf p})$. We first extend
the method used by Mattis and Lieb in 1D [7] to higher dimensions.

A fermion $\psi_s({\bf x})=\sum_{\bf p}e^{-i{\bf p}\cdot{\bf x}}c_{{\bf p}s}$
is written in terms of operators with momenta restricted to
each sector, $\Lambda_\alpha$, of the polygonised Fermi surface, {\it i.e.}
 $\psi_s({\bf x})
\approx \sum_{\alpha}\psi_{\alpha s}({\bf x})$
where $\psi_{\alpha s}({\bf p})
=\sum_{{\bf p}\in \Lambda_\alpha}e^{-i{\bf p}\cdot{\bf x}}c
_{\alpha,{\bf p},s}$.
Then the fermion momentum distribution at zero temperature is
$$
n({\bf p})
=\sum_\alpha\int d^2{\bf x}e^{i{\bf p}\cdot{\bf x}}<G|\psi_{\alpha s}
^\dagger({\bf x})\psi_{\alpha s}(0)|G>,
\eqno(12)
$$
where $|G>$ is the vacuum state of the interacting system described by the
bosonized Hamiltonian (8).

The ground state, $|G>$,
can be related to the non-interacting vacuum state $|0>$ by a canonical
transformation of the type $|G>=e^{-i\Sigma}|0>$, as pointed out by
Mattis and Lieb.
Under this transformation the fermion
operator becomes $\tilde{\psi}_s({\bf x})
=e^{-i\Sigma}\psi_s({\bf x}) e^{i\Sigma}$.
We assume
$\Sigma$ has the form
$i\sum_{n,D_n,\alpha, \beta} \Gamma^\dagger_{\alpha{\bf q}_n}
A_{\alpha\beta,{\bf q}_n}\Gamma_{\beta{\bf q}_n}$. The matrix $A$
can then be determined from $e^O=X$  with $O_{\alpha\beta}=\Delta_\alpha
A_{\alpha\beta}$.
As in the 1D case, the fermion operator in the bosonic representaion
is of the form $\psi_{\alpha s}({\bf x})=F_{\alpha s}({\bf x})e^{J_
{\alpha s}({\bf x})}$ where $F({\bf x})\sim e^{i{\bf p}_F^{(\alpha)}
\cdot{\bf x}}$ can be an arbitrary function of ${\bf x}$. $J({\bf x})$ is
the current operator and is very analogous to
its counterpart in 1D. Dividing $J$
into $J^\sigma$ and $J^\rho$ relating to spin and charge  currents
respectively and solving the `equation of motion' for $f_{\alpha\tau}=
e^{i\tau\Sigma}\psi_\alpha({\bf x})e^{-i\tau\Sigma}$ with time
`$\tau$' [7,5], one obtains
$$
\tilde{\psi}_{\alpha s}({\bf x})=W_\alpha({\bf x})Y_\alpha({\bf x})
\psi_{\alpha s}.\eqno(13)
$$
Since the elements of the transformation matrix $X$ satisfy
$X_{\alpha\beta}=X_{2N+\alpha,2N+\beta}$, the operators
$W_\alpha$ and $Y_\alpha$, in the cases for $\alpha=i=0,...,N-1$,
are given by
$$\begin{array}{rcl}\disp
W_i({\bf x})&=&
\disp {\rm exp}\frac{2\pi}{\sqrt{2\Lambda\Omega}}\{\sum_{D_\pm}
\frac{X_{ii}-1}{\omega^{1/2}_{i{\bf q}}}[e^{\pm i{\bf q}\cdot{\bf x}}
R_{i{\bf q}}-e^{\mp i{\bf q}\cdot{\bf x}}R^\dagger_{i{\bf q}}]\}\\[4mm]
Y_i({\bf x})&=&\disp {\rm exp}\frac{2\pi}
{\sqrt{2\Lambda\Omega}}\sum_{\beta\not=i}\{
\sum_{D_\pm}\frac{X_{i\beta}}{\omega^{1/2}_{i{\bf q}}}
[e^{\pm i{\bf q}\cdot{\bf x}}\Gamma_{\beta{\bf q}}
-e^{\mp i{\bf q}\cdot{\bf x}}\Gamma^\dagger_{\beta{\bf q}}]\}
\end{array}   \eqno(14)
$$
where $D_+=\{D_0,...,D_{N-1+i}\}$ and $D_-=\{D_{N+i},...,D_{2N-1}\}$.
Similar relations follow for $\alpha=N,...,4N-1$.
These results, combined
with the orthogonality relations (11), lead, after some algebra
similar to that in ref.[5], to
$$
<G|\psi^\dagger_{\alpha s}({\bf x})\psi_
{\alpha s}(0)|G>={\rm exp}\{-Q_\alpha({\bf x})\}
<0|\psi^\dagger_{\alpha s}({\bf x})\psi_{\alpha s}(0)|0>,\eqno(15)
$$
where
$$ \begin{array}{rcl}\displaystyle
Q_\alpha({\bf x})&=&
\disp\frac{(2\pi)^2}{\Lambda\Omega}\sum_{\beta,{\bf q}\in D_\beta}
\frac{1}{
\omega_{\beta{\bf q}}}f_\alpha({\bf q})({\rm cos}{\bf q}
\cdot {\bf x}-1),\\[3mm]
f_\alpha({\bf q})&=&\left\{\begin{array}{ll}
                   \disp \sum_{j=0}^{N-1}(X^2_{\alpha j}({\bf q})
                    +X^2_{\alpha,3N+j}({\bf q}))-1,
                    &\hbox{if $\lambda_\alpha>0$,}\\
\disp\sum_{j=0}^{N-1}(X^2_{\alpha,N+j}({\bf q})+X^2_{\alpha,2N+j}
       ({\bf q}))-1,
                    &\hbox{if $\lambda_\alpha<0$.}\end{array}
                    \right.
\end{array}\eqno(16)
$$

It is seen from (12) and (15) that the momentum distribution $n({\bf p})$
is given in terms of the quantities $Q_\alpha({\bf x})$.
Once these are known we can establish the
nature of any singularity in $n({\bf p})$ near $p_F$.
%The quantities, $X_{\alpha\mu}$, need to be found for each special form
%for the interaction.
We note that, although the unobservable
quantity $f_\alpha({\bf q})$ depends on the ordering of
$\{\lambda_\alpha\}$ via (16), the physical quantity $n({\bf p})$
does not.

To illustrate the method, we have solved for the
$X_{\alpha\mu}$ for the case of a simple $\delta$-function interaction.
In this case  $V({\bf q})=U$ and $V'({\bf q})=0$ and we can define the
dimensionless coupling constant
$
g=\pi p_FU/v.
$
We denote the component of ${\bf q}$  perpendicular
(parallel) to ${\bf n}_{\alpha_0}$
for any $\alpha_0$ as $q_\perp$ ($q_\parallel$) and scale them by
$t_\parallel=q_\parallel/L$ and $t_\perp=q_\perp/\Lambda$.
For any given $M$, $g$ and ${\bf t}$, we denote
by $f_{max}({\bf t})$ and $Q_{max}({\bf x})$ the values of
$f_{\alpha}({\bf t})$ and $Q_{\alpha}({\bf x})$ corresponding
to the maximum and minimum values of $\lambda_\alpha$,
$\pm \lambda_{max}$. We also order the $\lambda_\alpha$ so
that $\lambda_\alpha = \lambda_{max}$ for $\alpha=\alpha_0$
and $\lambda_\alpha = -\lambda_{max}$ for $\alpha=\alpha_0+2N$.

We start with some empirical observations on the results we obtain.
For small $g$ ($g\leq 0.05$) we find that
$f_{max}({\bf t})$ depends on ${\bf t}$ and is comparable to the other
$f_\alpha$.
However,
above $g\sim 0.05$ there is a crossover to a regime in which $f_{max}
({\bf t})$ does not depend on ${\bf t}$ and is significantly larger than
the other $f_\alpha$: $f_\alpha/f_{max}\sim 10^{-1}$ for $g=0.05$
while $f_\alpha/f_{max}\sim 10^{-3}$ for $g\geq 1.0$.
As $f_{max}$ turns out to correspond
to an exponent we denote it by $\delta$. We find that, for any given $g$,
$\delta$ converges to a constant as $M$ increases.
This is shown in Fig.\ 1.
When $\delta$ is much larger than the other $f_\alpha$,
we retain only $Q_{max}$ and set all other $Q_\alpha$ to zero.
This allows us to write
$$\begin{array}{rcl}
n({\bf p})&\approx&\disp\int d^2{\bf x} \int_{{\cal D}_{\alpha_0}}d^2{\bf k}
e^{-Q_{max}({\bf x})}[e^{-i({\bf k}-{\bf p})\cdot{\bf x}}+
e^{i({\bf k}+{\bf p})\cdot{\bf x}}]\theta(|{\bf k}|-p_F)\\[3mm]
{}~&+&\disp\sum_{\alpha\not=\alpha_0,\alpha_0+2N}\int_{{\cal
D}_\alpha}d^2{\bf k}\delta
({\bf p-k})\theta(|{\bf k}|-p_F), \end{array}\eqno(17)
$$
where the integration domains,
${\cal D}_\alpha$, are the sectors corresponding to
$D_\alpha$ in the full ${\bf k}$-space and $<0|\psi^\dagger({\bf x})
\psi(0)|0>=1/\Omega\sum_{\bf k}\theta(|{\bf k}|-p_F)$, the free fermion
distribution, is used.

We now evaluate $n({\bf p})$ from (17).
Since the angle
$\alpha_0$ can be chosen arbitrarily, it is convenient to assign $\alpha_0$
in such a way that it labels the sector containing ${\bf p}$.
The second term in (17) then vanishes.
Since ${\bf p}\in {\cal D}_{\alpha_0}$,
$Q_{max}({\bf x})\approx Q_{max}(x_\parallel,0)$ and the integration
over ${\bf t}$ leaves
$$
Q_{max}(x_\parallel,0)=\delta\int_0^{W}\frac{ds_\parallel}{s_\parallel}
(1-{\rm cos}~s_\parallel),     \eqno(18)
$$
where $W=Lx_\parallel/2$. For $W\gg 1$, this gives
$Q(x_\parallel,0)=\delta~ ln~W^2$ (see [5]) and, after integrating over
$x_\parallel$ in (18), we obtain
$$
n({\bf p})={\rm const.}-c\disp
(\frac{|{\bf p}|-p_F}{L}\disp )^{\delta}
{\rm sign}(|{\bf p}|-p_F), \eqno(19)
$$
with $c=2^\delta I(\delta)/\pi \delta$ and $I(\delta)
=\int_0^\infty ds_\parallel s_\parallel^{-\delta}~{\rm cos}~s_\parallel$.
The power exponent $f_{max}=\delta$
is the function of $g$, shown in Fig.\ 2.

For the case of intermediate coupling $g\geq 0.05$, where we can obtain the
approximate form for $n({\bf p})$ given in (19), one can see clearly
that the
2D system is not a Fermi liquid. In Fig.\ 2 we compare the
variation of $\delta$ with $g$ to the function $\delta(g_1)
=\disp\frac{1+g_1}{\sqrt{1+2g_1}}-1$ for
the 1D case ($g_1=2U/v$).
For the 2D case the non-Fermi-liquid
exponent is always smaller than in the 1D case for comparable values
of the dimensionless coupling constants.

For weak coupling $g\leq 0.05$, the contributions from $\pm\lambda_{max}$
are not dominant and contributions from all of $\lambda_\alpha$ are
comparable. We have noticed that the largest of the $f_\alpha$
varies as $10^{-2(n+1)}$ for $g\sim 10^{-n}$. This clearly implies that it
will be difficult to distinguish between Fermi-liquid and
non-Fermi-liquid-like behaviors. This might be the reason why the
non-Fermi-liquid state has not been found for weak interactions using
perturbative approaches [3].

We would like to comment on the neglect of the around-the-corner(ATC)
scattering. One might expect that in the limit of a large number of sectors
($\Lambda \to 0$), ATC scattering should become increasingly important, as
more and more states are close to the corner of a sector. However, as our
results show that the exponent, $\delta$, is already well-converged
for a small number of sectors, we do not believe that neglect of ATC
scattering poses serious problems.

Although we have located at
 the bosonization of the Hamiltonian for a homogeneous two-dimensional
system, we end up with a Hamiltonian which resembles that for coupled
chains discribed by Wen ${[10]}$. Our results differ from those he
obtained for coupled chains in one very significant feature. In Wen's
case the interchain interactions modify the single chain Luttinger liquid
state. In our case the interactions between sectors generate the
non-Fermi liquid state. If we switch off this interaction the
system reverts back to a Fermi liquid state and not a series of uncoupled
1D Luttinger liquids.

In conclusion, we have discussed a generalization to 2D
of the Tomonaga-Luttinger model using Haldane's bosonization procedure.
The bosonized system has been solved exactly by a generalized Bogoliubov
transformation. The fermion momentum distribution at zero temperature
has been
calculated using an extension of the Mattis-Lieb technique. We have found that
the momentum distribution exhibits a non-universal power law behavior in the
vicinity of the Fermi surface for intermediate interaction strengths.
It seems to us that this is the first time that Luttinger
liquid behavior has been found nonperturbatively
for a two-dimensional model with a non-singular interaction.

We are very grateful to D. P. Chu for useful discussions.
This work was supported in part by SERC of the United Kingdom under
grant No.GR/E/79798 and also by MURST/British Council under grant
No.Rom/889/92/47.

\eject

\bigskip
\begin{flushleft}
{\bf REFERENCES}
\end{flushleft}
\medskip
\begin{description}
\item{[1]} P. W. Anderson, Phys. Rev. Lett. {\bf 64}, 1839 (1990);
{\bf 65}, 2306 (1990); {\bf 66}, 3226 (1991);{\bf 67 }, 2092 (1992).
\item{[2]} F. D. M. Haldane, J. Phys. {\bf C 14}, 2585 (1981).
\item{[3]} J. R. Engelbrecht and M. Randeria, Phys. Rev. Lett.
{\bf 65}, 1032 (1990); {\bf 66}, 3325 (1991); Phys. Rev. {\bf B 45},
12419(1992); J. R. Engelbrecht, M. Randeria and L. Zhang,
Phys. Rev. {\bf B 45}, 10135 (1992); P. A. Bares and X. G. Wen
preprint, 1992; C. Castellani, C. Di Castro and W. Metzner, preprint,
1993. R. Shankar, to be published.
\item{[4]} M. Fabrisio, A. Parola and E. Tosatti, Phys. Rev. {\bf
B 44}, 1033 (1991).
\item{[5]} D. V. Khveshchenko, R. Hlubina and T. M. Rice,
preprint, ETH-TH/93-3, 1993, to appear in Phys.Rev. {\bf B}.
\item{[6]} F. D. M.  Haldane, Varenna lectures, 1992 and Helv. Phys.
Acta. {\bf 65}, 152 (1992); also see A. Houghton and B. Marston,
Brown University Preprint (1992).
\item{[7]} D. C. Mattis and E. Lieb, J. Math. Phys. {\bf 6} ,
304 (1965).
\item{[8]} See, for example, G. D. Mahan, {\it Many-Particle Physics},
Plenum Press, New York and London, (1990, 1981).
\item{[9]} We have checked that for $\delta$-function interaction
all of $\lambda_\mu$ are real for large $M$.
\item{[10]} X. G. Wen, Phys. Rev. {\bf B 42},6623 (1990).
\end{description}
\eject
\begin{flushleft}
{\bf CAPTIONS OF FIGURES}
\end{flushleft}
\begin{description}
\item{Figure 1}.  The exponents $\delta$ as a function of the number
of sectors of the polygonised Fermi surface, $M$, for various values of
the dimensionless coupling constant $g$. The exponent, $\delta$,
characterises the singularity at
the Fermi surface in the momentum distribution function, $n({\bf
p})$. As $M$ increases, $\delta$ converges to
a constant value for each coupling constant $g$.
\item{Figure 2}.  The $M\rightarrow \infty$ limit for the exponent,
$\delta$, as a function of coupling constant
$g$. ($M$ is the
number of sectors of the polygonized Fermi surface). We also show the
corresponding results for the 1D case.
\end{description}
\end{document}